\documentclass{aastex62}

\graphicspath{{./}{figures/}}

\received{April 1, 2019}
\revised{March 5, 2021}
\accepted{\today}
\submitjournal{ApJ}

\shorttitle{Observations of Thomson scattering from a loop-prominence system}
\shortauthors{Mart\'inez Oliveros et al.}

\begin{document}

\title{Observations of Thomson scattering from a loop-prominence system}

\correspondingauthor{Juan Carlos Mart\'inez Oliveros}
\email{oliveros@ssl.berkeley.edu}

\author[0000-0002-2587-1342]{Juan Carlos Mart\'inez Oliveros}
\affil{Space Sciences Laboratory, University of California, Berkeley, CA 94720-7450, USA}

\author[0000-0001-6854-2779]{Juan Camilo Guevara G\'omez}
\affiliation{Rosseland Centre for Solar Physics, University of Oslo, Postboks 1029 Blindern, 0315 Oslo, Norway}
\affiliation{Institute of Theoretical Astrophysics, University of Oslo, Postboks 1029 Blindern, 0315 Oslo, Norway}

\author[0000-0002-8283-4556]{Pascal Saint-Hilaire}
\affil{Space Sciences Laboratory, University of California, Berkeley, CA 94720-7450, USA}

\author[0000-0001-5685-1283]{Hugh Hudson}
\affil{Space Sciences Laboratory, University of California, Berkeley, CA 94720-7450, USA}
\affil{SUPA Department of Physics and Astronomy, University of Glasgow, G12 8QQ, UK}

\author[0000-0002-2002-9180]{S\"am Krucker}
\affil{Space Sciences Laboratory, University of California, Berkeley, CA 94720-7450, USA}
\affiliation{Institute for Data Science, School of Engineering, University of Applied Sciences and Arts Northwestern Switzerland, 5210 Windisch, Switzerland}

\begin{abstract}
We describe observations of the white-light structures in the low corona following the X8.2 flare SOL2017-09-10, as observed in full Stokes parameters by the Helioseismic and Magnetic Imager (HMI) of the Solar Dynamics Observatory.
These data show both bright loops and a diffuse emission region above them.
We interpret the loops as the white-light counterpart of  a classical loop-prominence system, intermediate between the hot X-ray loops and coronal rain.
The diffuse emission external to the loops is linearly polarized and has a natural interpretation in terms of Thomson scattering from the hot plasma seen prior to its cooling and recombination.
The polarimetric data from HMI enable us to distinguish this contribution of scattering from the HMI pseudo-continuum measurement, and to make a direct estimation of the coronal mass in the polarized source. For a snapshot at 16:19~UT, we estimate a mass $8 \times 10^{14}$~g.
We further conclude that the volumetric filling factor of this source is near unity.

\end{abstract}


\section{Introduction}
Classical loop-prominence systems { associated with powerful solar flares} have been observed since the 1930's \citep[e.g.][]{1964ApJ...140..746B,1968IAUS...35..287S}. These are commonly observed in H$\alpha$ and other chromospheric lines  in the form of loop arcades. Since their initial observations, loop prominence systems (LPS) were a topic of particular interest for solar physicists as they seemed to have a clear relationship with solar flares. 
Moreover, LPS display filamentary structures with episodes of plasma condensation \citep{1963ZA.....56..291W}, leading to the 
somewhat misleading term ``sporadic coronal condensation'' for the flare-associated LPS.
\citet{1968IAUS...35..287S} found that nearly 70\% of all LPS are observed in active regions, an in particular active regions that also hosted solar flares
and were often associated with the coronal yellow line of Ca~{\sc xv}, implying higher than normal active-region temperatures.
The early observations further suggested that the LPS are highly correlated with active regions, and in particular with active regions that can produce highly energetic flares (e.g.  proton flares) \citep{1963ZA.....56..291W}. 
Since their discovery loop-prominence systems have been identified differently with ``loop-prominences", ``flare arcades", ``eruptive loops", ``post-flare loops", or ``post-eruptive arcades'' \citep[for an in-depth discussion see][]{2007SoPh..246..393S}. 
Although similar in nature, loop-prominence systems  should not be confused with ordinary prominences. 
According to \citet{2007SoPh..246..393S} the key difference between these two phenomena is their evolution. 
Prominences grow by expanding the individual loops that compose them, while loop-prominence systems grow by the illumination of new higher loops while the lower fade. 
The time evolution of loop prominence systems is related to the thermodynamic and radiative properties of the  chromospheric-temperature plasma. From observations \citep[e.g., ][]{1964ApJ...140..746B} it is clear that these systems do not grow from lower to higher altitudes \citep[see Figure 2 in ][]{1963ZA.....56..291W}, instead their generation is the product of higher loops getting more intense due to cooling to low temperatures, while the lower ones decay. 
This evolutionary process can take several tens of minutes to several hours \citep{2007SoPh..246..393S}. 
We now understand this in terms of the cooling of arcade loop systems from X-ray temperatures, down through Ca~{\sc xv} and then abrupt condensation to produce emission in cool lines like H$\alpha$.

Although LPS commonly occur in flares near the limb, their observation in white light or pseudo-white light above the solar limb is extremely rare, even more so without the usage of coronagraphs since the signatures are faint relative to the intensity of the emission from the disk.
Only a few cases exist in the literature, to our knowledge,  describing visual observations of such events: SOL1980-06-21 (X2.6), an event visually observed by J. Harvey and T. Duvall \citep{1983SoPh...86..123H,1990ApJS...73..213C}, SOL1989-08-14 \citep[$\sim$X20; ][]{1992PASJ...44...55H}, and SOL2003-11-04  \citep[$>$X17;][]{2004AAS...204.0213L} may comprise the whole list of such visual
observations.
Recently, though, \cite{2014ApJ...780L..28M} and \cite{2014ApJ...786L..19S} found two excellent examples in ordinary optical imaging, but from space: 
SOL2013-05-13T02:17 and SOL2013-05-13T16:00.
These observations made use of the Helioseismic and Magnetic Imager \citep[HMI; ][]{2012SoPh..275..229S,2012SoPh..275..207S} of the Solar Dynamics Observatory \citep[SDO; ][]{2012SoPh..275....3P}.
\cite{2018ApJ...867..134J} have now also analyzed SOL2017-09-10 using HMI data, and in this paper we analyze its polarization properties.

Our study of SOL2017-09-10 focuses on the development of the thermal sources in the flare.
Many earlier papers describe this fascinating event in great detail, but mainly bypassing the basic thermal emissions in favor of studying more complicated aspects of the flare development, such as current-sheet formation \citep[e.g.][]{2018ApJ...854..122W} and magnetic reconnection, especially via the new capabilities of microwave imaging spectroscopy \citep[e.g.][]{2018ApJ...863...83G,2020ApJ...900...17Y,2021ApJ...908L..55C}.

The HMI observations describe off-limb emission at the solar limb and relatively low in altitude (below 1.03 R\textsubscript{\(\odot\)} elongation). 
Both of the 2013 flares, the first reported from the HMI database, also showed white-light (WL) footpoint sources at the level of the photosphere. 
The gradual coronal emissions can be identified as visual counterparts of the classical loop-prominence system, but were brighter than expected and have some ambiguity in their emission mechanisms; a continuum could come from scattering or from the free-bound extension of the hot plasma seen in soft X-rays, or from direct emission (lines and continua), depending mainly upon the density \citep{2018ApJ...867..134J}.
In their interpretation, the coronal sources detected by HMI in these flares represent flare loops, initially heated to X-ray temperatures, and detected in the process of cooling. 
The authors found the HMI flux to exceed the long-wavelength extrapolation of the bremsstrahlung of the flare soft X-ray sources by at least one order of magnitude, implying the contribution of cooler material that could produce free–bound continua and possibly line emission detectable by HMI.
Their analysis suggested electron densities as high as 10$^{13}$~cm$^{-3}$, and under these conditions optically-thin Paschen continuum is not important.

\citet{2014ApJ...786L..19S} reported the detection of linearly polarized scattered light from an LPS (SOL2013-05-13T02:17 and SOL2013-05-13T16:00) via the HMI white-light polarimetric data. 
This revealed Thomson scattering and therefore enabled a direct mass estimate and inferences about the source density, using the techniques described in \citet{1930ZA......1..209M}. 
\citet{2014ApJ...786L..19S} concluded that only a fraction of white-light emission in LPS was due to Thomson scattering and inferred a lower limit of the free electron density of about 3.5$\times$10$^{11}$ cm$^{-3}$, assuming a line-of-sight depth of  2.2~Mm.

\begin{figure}[htbp]
\centering
\includegraphics[width=0.49\textwidth]{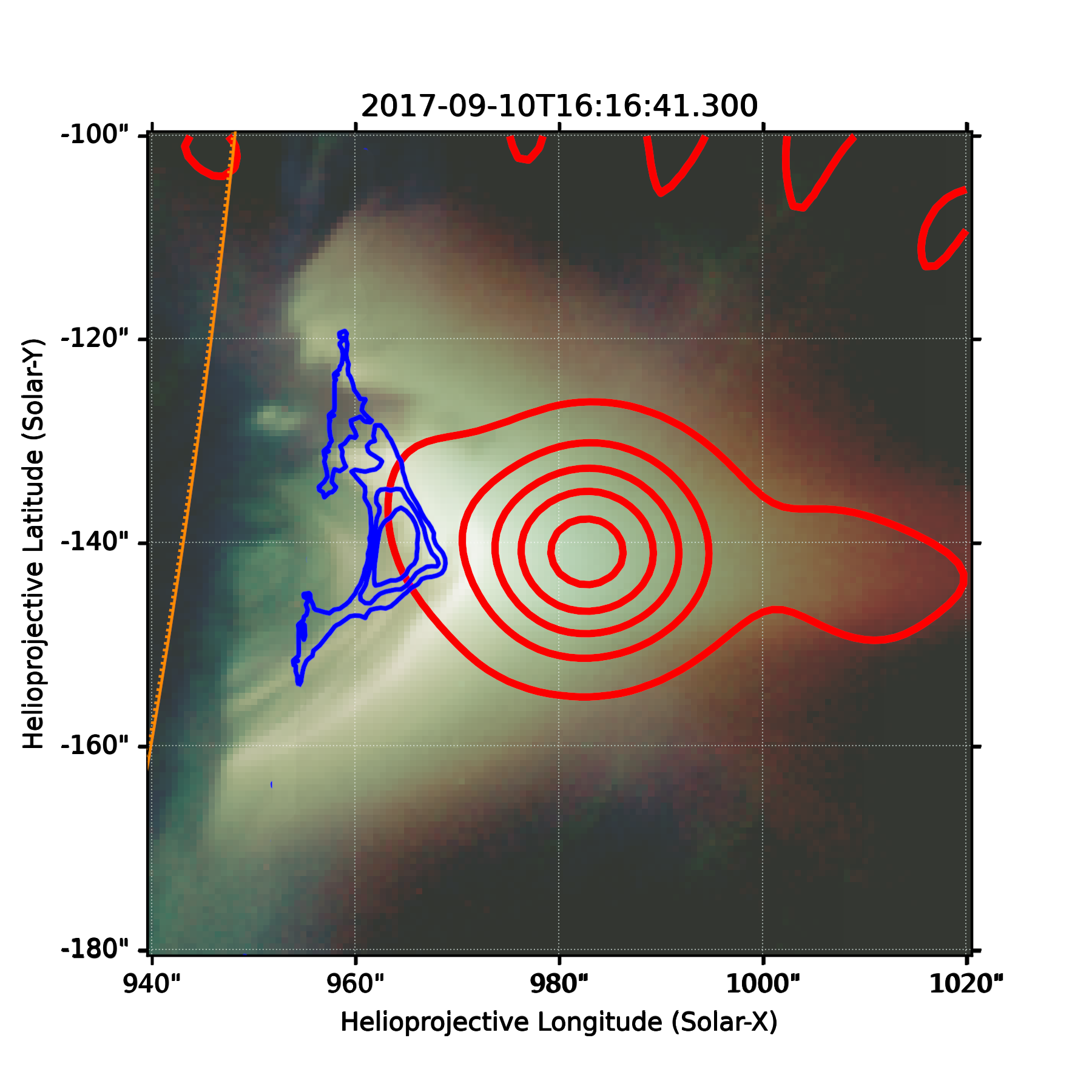}
          \caption{\textit{A snapshot of the multiwavelength sources observed in SOL2017-09-10 at 16:16:41~UT, with the background image an AIA 94/193/211 composite.
          The blue contours show the bright HMI loops at this moment, and 
          the RHESSI image at 12\,--\,18~keV is shown by red contours (10\%, 30\%, 50\%, 70\%, 90\%).} 
      }
\label{fig:smapaia_94_sswl}
\end{figure}

In this article, we describe similar polarization measurements and analysis of the well-studied SOL2017-09-10 limb flare. 
The HMI white-light intensity (i.e., Stokes I) of this event were reported by \citet{2018ApJ...867..134J}, whose study took into account all continuum emission processes in the flare loops, as well as optical-depth effects; see also \cite{1992PASJ...44...55H}.
The LPS appear in an image annulus extending about 44$''$ above the limb, evolving for several hours in a close relationship with the EUV structures observed by AIA \citep[e.g.][]{2018ApJ...854..122W,2020ApJ...896L..35H}.
For reference, we show the spatial relationships of these sources in Figure~\ref{fig:smapaia_94_sswl}.
The diffuse HMI source above the LPS was bright enough, in this case,
for us to make a first estimate of the density/filling factor for the hot plasma routinely seen as recombination radiation in the soft X-ray and microwave bands in major flares.
We note that \cite{2019ApJ...887L..34F,2020ApJ...900..192F}, \cite{2021ApJ...921L..26Z} and \cite{2018ApJ...866...64C} have also analyzed this event with the complementary Mauna Loa K-Coronameter observations.

\section{Observations and Instrumentation}

On 2017 September 10, a loop prominence system developed during the gradual phase of an X8.2 flare (SOL2017-09-10). 
This LPS was well observed by the SDO/AIA \citep[e.g.][]{2018ApJ...854..122W} and various other aspects of this classical eruptive flare event have been widely reported in the literature.
In particular, \cite{2018ApJ...867..134J} have previously reported on the HMI observations, but not using its polarimetric capability.

The HMI instrument contains two cameras (front and side) designed for on-disk observations of the whole Sun, primarily for helioseismology and the characterization of the photospheric magnetic field, and for these purposes it makes high-resolution filtergrams of the photospheric Fe {\sc i} absorption line at 6173.3~\AA. 
HMI observes the line via  six passbands spanning a range of about 345~m\AA~around the target line.
It also obtains full Stokes profiles \citep{2012SoPh..275..229S}. The ``front camera" standard data are created from a combination of sequences of 12 distinct images (six filters and two polarizations each called a filtergram), while the ``side camera" data are  reconstructed from observations with the same filters and the following polarizations: I+/--U, I+/--Q, and I+/--V. The individual wavelengths (and polarization settings) for these line profiles are not observed simultaneously, but in a programmed sequence extending over 45~s and 135~s frames for the HMI front and side cameras, respectively,  and therefore have limitations when rapid transients occur \citep[e.g.,][]{2011SoPh..269..269M,2014SoPh..289..809M}. The resulting images are cropped and later during the process, truncated at about 65$''$ above the limb to generate the standard HMI observables (intensity, velocity and magnetic field). The observations we report here, though unambiguous photometrically, thus occur in a parameter space not optimized by the design of the telescope.

\begin{figure}[!htb]
    \centering
    \includegraphics[width=0.75\linewidth]{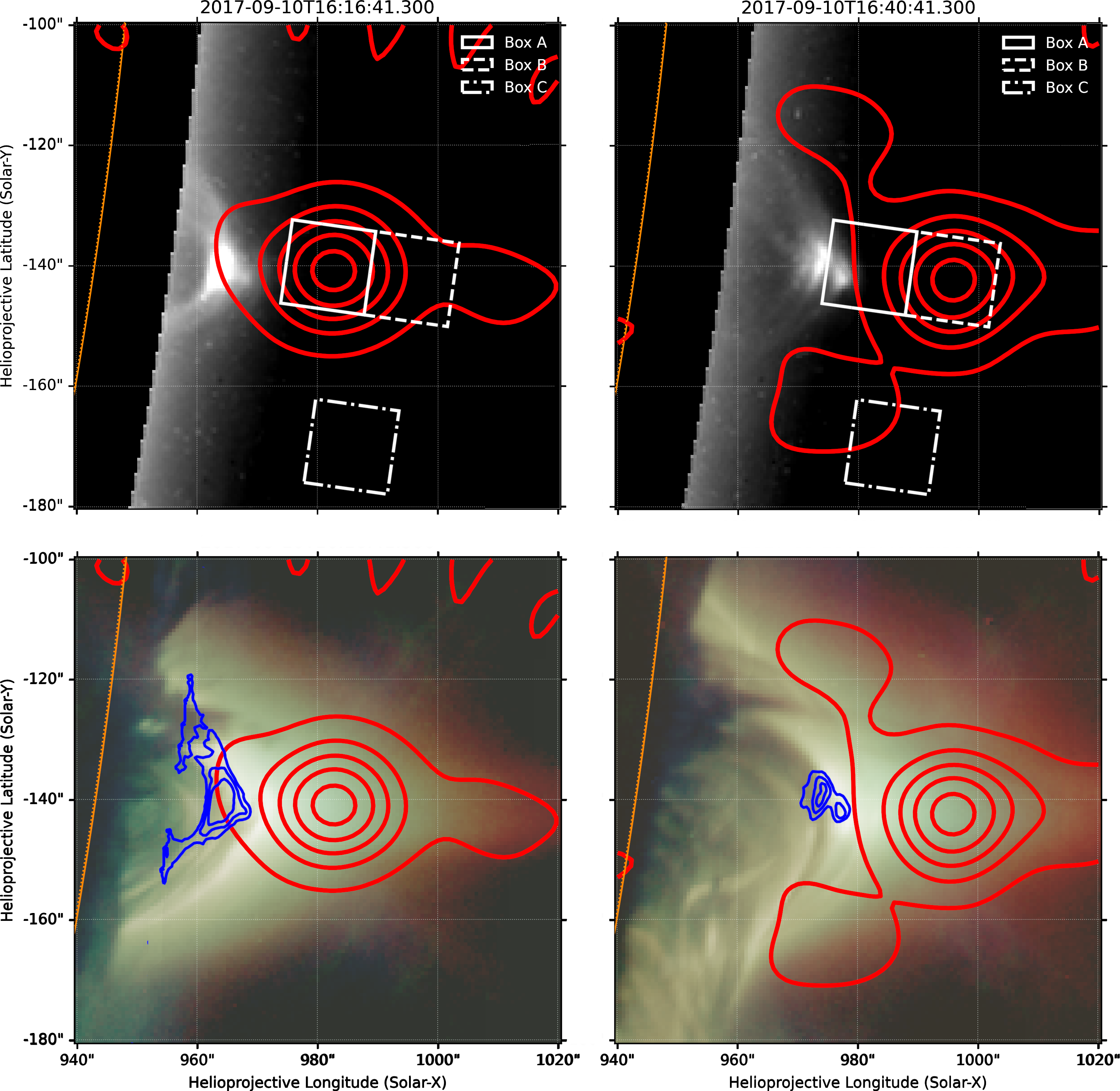}
    \caption{Top: HMI/Stokes I at two different times during the evolution of the LPS. White boxes A, B and C show the regions of analysis. Bottom: AIA 94/193/211 combination at same times, showing presence of cool plasma; blue contours are for the LPS feature from HMI/Stokes I at  75\%, 85\%, 95\%. The RHESSI 12\,--\,18~keV intensity contours at 10\%, 30\%, 50\%, 70\%, 90\% are shown in all the images in red color for two evolutionary times close to the HMI image time.
    The lower-left panel matches Figure~\ref{fig:smapaia_94_sswl}.
    The orange solid line represent the SDO/HMI limb}
    \label{HMI-AIA}
\end{figure}

Figure~\ref{HMI-AIA} shows the evolution of the LPS at two different and characteristic times, for HMI/STOKES I (top row) and AIA 94/193/211~\AA~composite images (bottom row). 
The RHESSI 10\%, 30\%, 50\%, 70\%, 90\% intensity contours are shown in red  and plotted for the same times. The RHESSI images were reconstructed using the CLEAN algorithm with a period of integration that encompasses the HMI image cadence. 
Figure~\ref{HMI-AIA} (bottom row) shows the HMI/STOKES I contours (blue) for each evolutionary time at 75\%, 85\%, 95\% over the composite AIA images. 
On the top row three regions (white boxes) are shown and designated as A, B and C. Regions A and B denote the areas where the HMI/Stokes Q, U and V were analyzed. Region C is our control region.

The data provided by HMI team through the Joint Science Operations Center (JSOC) has a cadence of 90 seconds and a roll angle or rotation angle  close to 180$^\circ$. The analysis to be described was made before applying the roll angle rotation in order to avoid mathematical artifacts introduced by the interpolation. The data also have a small jitter in every frame, so, the first operation was to recenter all frames in order to have all the images properly aligned in time and later the six filters were average for each Stokes component. In addition, to increase the signal,  images were summed over four consecutive frames to produce a single image every 6 minutes, which gives us sufficient temporal resolution for the purpose of this research. Finally, Q and U were rotated to increase the signal to noise ratio for the Stokes parameters throughout by aligning the plane of reference of the measurement with the location of the LPS \citep[see ][]{2004JGRD..109.9205S}. These new Stokes are named Q$'$ and U$'$ on the expectation that most of the polarized signal, in the case of Thomson scattering above a uniform photosphere, will appear in Q$'$ and little in U$'$ \citep{2014ApJ...786L..19S,2021ApJ...923..276S}.

\begin{figure}[!htb]
    \centering
    \includegraphics[width=0.75\linewidth]{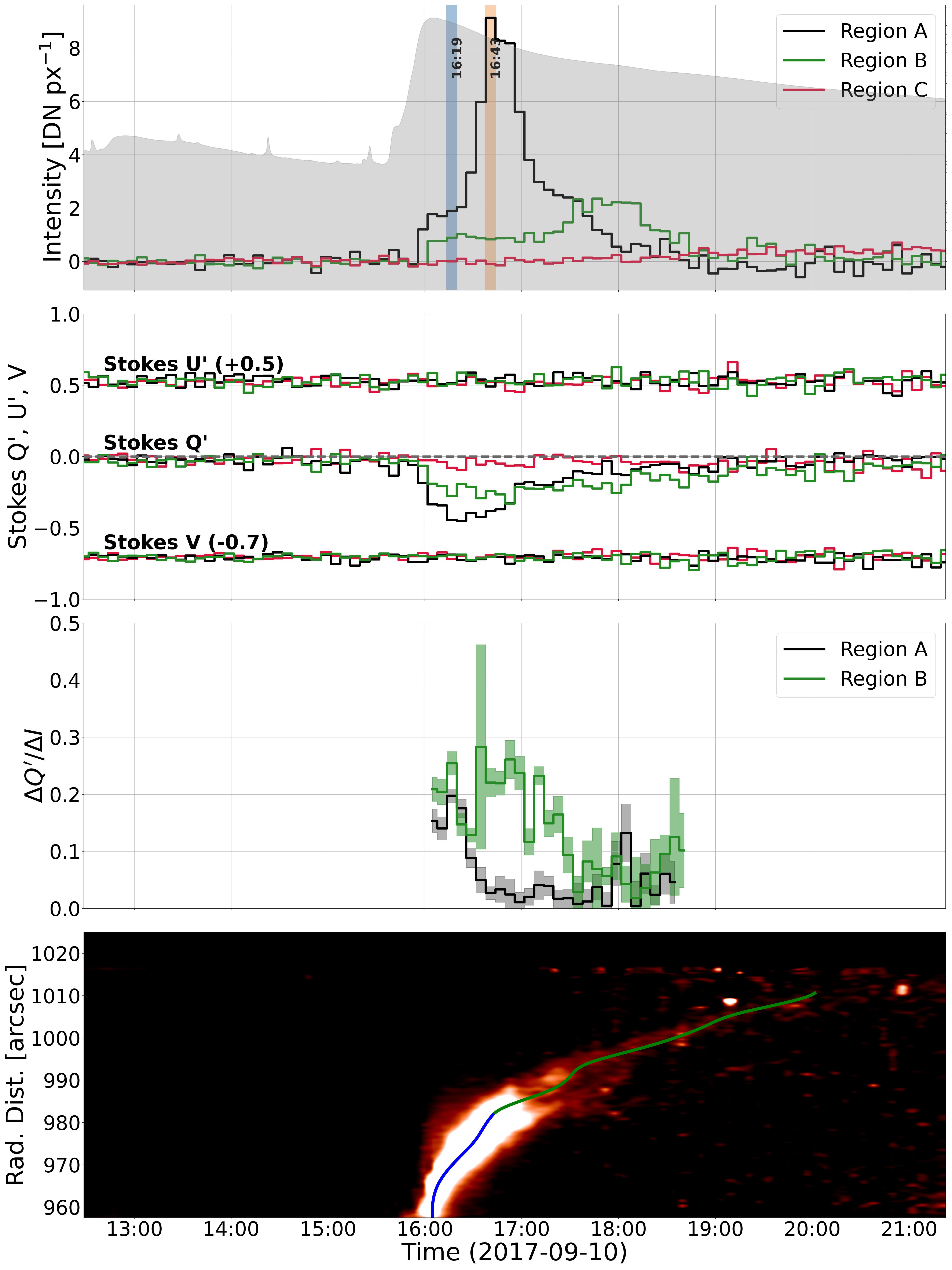}
    \caption{SDO/HMI Stokes parameters in the regions A, B, C. The top panel shows the intensity values as a function of time with GOES 1-8\AA~in gray. The second panel from the top shows the Q$'$, U$'$ and V Stokes parameters. The third panel shows the ratio $\Delta Q'/\Delta I$ for the regions A and B. The bottom panel shows the spatial and temporal evolution of the LPS intensity within a 10$''$ slit centered at the LPS brightest feature, the blue line indicates the apparent motion of the LPS with an average velocity of $7.88~\text{km\,s}^{-1}$ whereas the green line indicates the final phase of the event with an average velocity of $1.75~\text{km\,s}^{-1}$.}
    \label{IQUV_tdist}
\end{figure}

Figure~\ref{IQUV_tdist} shows the measured SDO/HMI Stokes parameters in the three regions of interest (A, B, C). The top panel shows the intensity values as a function of time. The GOES 1-8~\AA~X-ray flux is shown as the gray shadow area. It is clear that the highest value is reached in region A which during the event contains the visible part of the LPS. Region B is not as intense as region A and its peak occurred several tens minutes later than the peak in region A.  Our control region C shows no appreciable changes in intensity that can be attribute to the LPS. 
The second panel from the top shows the Stokes parameter Q$'$, U$'$ and V. Regions A and B show a change in the Stokes parameter (Q$'$) during the event of about 20\% with a maximum close to 44\%. 
The control region C shows no apparent change. 
The U$'$ and V Stokes parameters show values close to zero. 
The third panel shows the  $\Delta Q’$/$\Delta I$ values for regions A and B, region C is not shown since the values are undetermined in the time interval. Here we note a 19\% increase of this ratio in region A and about 21\% in region B in the time intervals selected to match the RHESSI observations. The bottom panel shows the spatial-temporal evolution of the LPS intensity within a 10$''$ slit centered at the LPS brightest feature. We calculated the average propagation velocity of the LPS during two stages of its evolution. The blue line indicates the apparent motion of the LPS with an average velocity of $7.88~\text{km\,s}^{-1}$ whereas the green line indicates the final phase of the event with an average velocity of $1.75~\text{km\,s}^{-1}$, these values are in agreement with those reported for this type of structures \citep[e.g.][]{1964ApJ...140..746B,2002SoPh..210..341G}. 

Table~\ref{tab:obs} summarizes the measurements of the excess fluxes for Boxes A, B and C at the two reference times indicated in Figure~\ref{HMI-AIA}, and interpreted as successive loops cooling and draining sequentially.

\begin{table}[h]
\centering
\caption{Observed Fluxes}
\medskip
\begin{tabular}{ l l l l l l }
\hline
Box & Time (UT) & Position & I (DN/s) & Q$'$ (DN/s) & Area (arcsec$^2$)\\
\hline
A & 16:19 & [982,-140] & 48.7$\pm1.5$ & $\sim$ -0.44$\pm$0.05 & 196 \\
B & 16:19 & [996,-142] & 39.9$\pm1.5$ & $\sim$ -0.27$\pm$0.03 & 196 \\
C & 16:19 & [996,-170] & 37.6$\pm1.5$ & $\sim$ -0.08$\pm$0.009 & 196 \\
\hline
A & 16:43 & [982,-140] & 56.1$\pm1.4$ & $\sim$ -0.38$\pm$0.04 & 196 \\
B & 16:43 & [996,-142] & 39.9$\pm1.4$ & $\sim$ -0.24$\pm$0.03 & 196 \\
C & 16:43 & [996,-170] & 37.5$\pm1.4$ & $\sim$ -0.03$\pm$0.004 & 196 \\
\hline
\end{tabular}
\label{tab:obs}
\end{table}

\section{Mass, density and filling factor from Thomson scattering}\label{sec:cd}

In the following sections we discuss the region \textit{above} the bright loops.
This faint and apparently diffuse volume corresponds to the hot plasma of the soft X-ray source, an almost definitive flare component and one that is not well imaged by EUV observations such as those of AIA.

\subsection{Mass}

The K-component of the white-light continuum observed from coronal structures consists of Thomson-scattered photospheric continuum, with the resulting linear polarization \citep[e.g.][]{1966gtsc.book.....B}.
This component of coronal brightness scales directly with $N_e$ (the column electon density) and hence the electron density $n_e$, while collisional atomic emissions (collisional bremstrahlung) scale as $n_e n_i$.
The total free electron number ${\mathcal N}_e$ then determines the mass of the scattering material,
if the abundances are known and if the location of the source justifies the ``plane-of-the-sky'' approximation.
At high densities the atomic emissions processes may dominate and even produce optically-thick continuum \citep{1992PASJ...44...55H, 2018ApJ...867..134J}.
This may obscure a part of the polarization source; accordingly
the mass estimate represents a lower limit.
Such an effect is likely to be a small one because of the efficiency of the thermal collapse \citep{1965ApJ...142..531F}.
The dust-scattered F-corona is not important in the lower corona.
Thus where Thomson scattering dominates, its brightness directly determines the mass of the scattering material, if the photospheric radiation field is known \citep{2021ApJ...923..276S}:
\begin{equation}\label{eq:scat_mass}
M = \mu m_p \mathcal{N}_e = \frac{F_s}{F_\lambda} \cdot \frac{\mu m_p (\pi  R_\odot^2) }{\sigma_T W} \ \ ,
\end{equation}
where $\mu$ is the mean molecular weight, $m_p$ the mass of the proton, $\mathcal{N}_e$ the number of scattering electrons, $F_s/F_\lambda$ the ratio of scattered irradiance to the total solar spectral irradiance at the HMI wavelength, $W$ a geometrical factor as estimated by \cite{2014ApJ...786L..19S}, and $\mathrm \sigma_T = 6.65\times10^{-25}~\text{cm}^{-2}$  the Thomson cross section.
This estimate assumes fully ionized plasma and does not depend upon the source density or filling factor $\eta$, which we discuss in the following sections.
Note that we use $\mathcal{N}_e$ for the total number of scattering electrons, $N_e$ for the column density (cm$^{-2}$), and $n_e$ for the number density (cm$^{-3}$) here and below.

\subsection{Density}\label{sec:density}

With further assumptions we can estimate the density of the scattering source and its filling factor. 
The basic idea of this technique is to derive the Thomson scattering contribution of the measured flux using the ratio between the observed linear polarization and the total intensity data. Following \citet{1994A&A...290..553R} and \citet{2009SoPh..254...89J} the density $n_e$ of an off-limb source emitting by Thomson scattering is:

\begin{equation}
    n_e = \frac{j(\nu)}{I_0(\nu)}\frac{1}{W}\frac{1}{\sigma_T} = \frac{\varepsilon_I}{I_0}\frac{1}{\sigma_T}\frac{1}{G_I} \, \text{\hspace{3mm} (eq. 14 in \citealt{2021ApJ...923..276S})}
    \label{density_1}
\end{equation}
where $j(\nu)$ is the volume emissivity at a given prominence location, I$_0(\nu)$ is the emitted intensity from the solar disk center, $W = G_I$ is a geometrical dilution factor dependent on the height over the limb which represents our knowledge of the source structure. \citet{2014ApJ...786L..19S} described how to use the theoretical values from \citet{1930ZA......1..209M} in order to rewrite Equation~\ref{density_1} in terms of the effective Thomson scattering cross section at a given height $H$ over the solar limb $\sigma_T(H) = \sigma_T \cdot G_I(H)$\footnote{$G_I(H)$ is $\approx$~0.3 for the observed source heights and the known limb darkening at 6173\AA}, the line-of-sight thickness $D$ and the ratio between the scattered intensity from the source at the given height $\Delta I(H)$, understood as the intensity excess value above the baseline, and the intensity from the solar disk center such that:

\begin{equation}\label{density_2}
    n_e(H) = \frac{\Delta I(H)}{I_0}\frac{1}{\sigma_T(H)}\frac{1}{D}\ \ .
\end{equation}
This allows us to estimate the electron density of a source, with $\Delta I$ being the intensity produced by only the scattered-light component. 
The observed $\Delta I$, however, also may have a component due to optically-thick sources seen in emission. 
As these latter sources are unpolarized and the expected degree of polarization due to Thomson scattering is known as a function of height \cite[e.g.][]{2014ApJ...786L..19S}, the two components can be separated using the actual measured degree of polarization $P_m$ and the expected one $P_T$. Taking this factor into account in Equation~\ref{density_2}, it is possible to calculate the density of the material responsible for the Thomson-scattered signal: 

\begin{equation}\label{density_3}
    n_e = \frac{\Delta I}{I_0}\frac{1}{\sigma_T(H)}\frac{P_m}{P_T}\frac{1}{D}\ \ .
\end{equation}
The column density $N$~(electrons cm$^{-2}$) is implicit in Equation~\ref{density_3}; it is the integral of the electron density along the line of sight through the emitting volume. 
Here we estimate the column depth by taking the width of the 70\% image contour of the RHESSI source.
This is shown as the box dimension in Figure~\ref{HMI-AIA}.
Thus, the column density, with the polarization-factor correction, becomes:

\begin{equation}\label{density_4}
    N_e = \frac{\Delta I}{I_0}\frac{1}{\sigma_T(H)}\frac{P_m}{P_T}\ \ .
\end{equation}

\begin{figure}[htbp]
    \centering
    \includegraphics[width=0.95\linewidth]{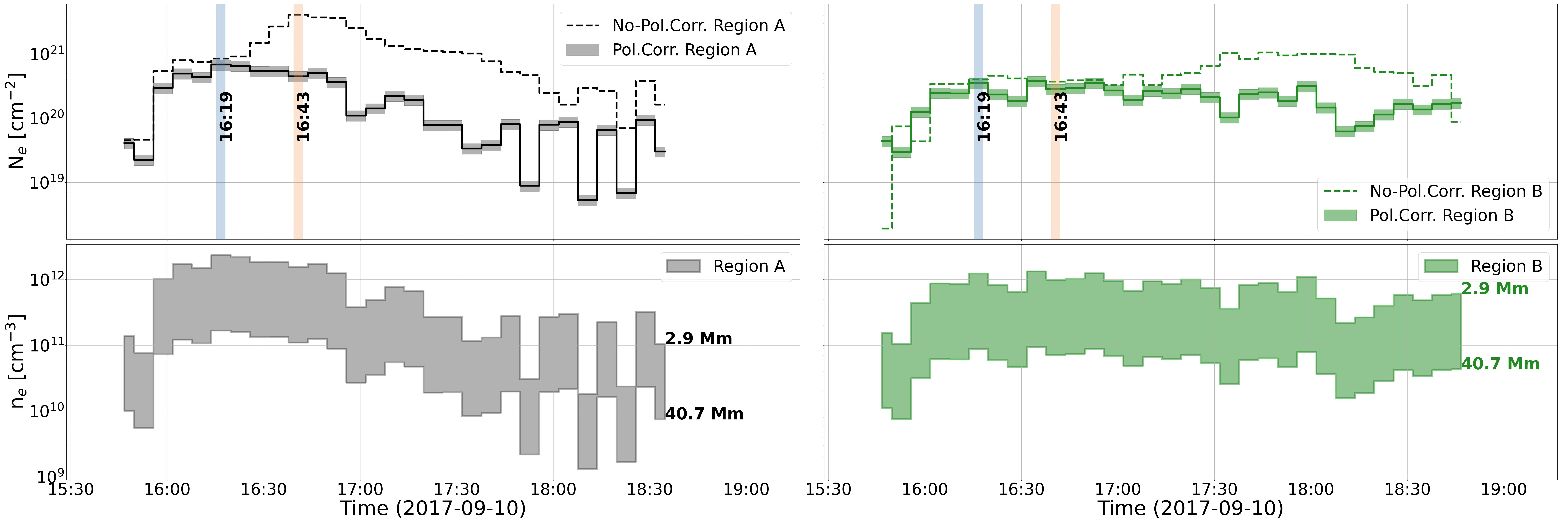}
    \caption{Top left and right panels show the column densities for regions A and B respectively, the shadow refers to the errors in each case. In both cases, the dashed curve is the column density without correction by the polarization factor whereas the solid and shadowed region have the correction done. The bottom panels are the densities for the regions A and B derived from the column densities assuming a \textit{line-of-sight} thickness between $2.9~\text{Mm}$ and $40.7~\text{Mm}$ (shadowed regions).}
    \label{fig:Ne_ne}
\end{figure}

The upper panels of Figure~\ref{fig:Ne_ne} 
show the column densities for regions A and B respectively, as calculated using Equation~\ref{density_4}. If the depolarization factor is not applied, i.e., if the flux originated entirely from Thomson scattering, the column densities would follow the dashed curves. 
The correction factor in each box increases with time,
as expected for sources cooling and recombining.
For region A, which is the closest to the photosphere and therefore is reached first by the white-light structure seen in HMI/Stokes~$I$ (see Figure~\ref{HMI-AIA}), the uncorrected column density is as large as $10^{21}~\text{cm}^{-2}$. However, when the column density is corrected by the polarization, gray curve in top left panel, it drops by almost an order of magnitude as this represents the actual Thomson scattered component only. For region B, there is no significant variation between the uncorrected and corrected column densities initially, meaning that the source is mainly dominated by Thomson scattering until the observed structure reaches it at a later stage. 
Finally, the bottom panels shows the expected electron densities derived from the corrected column densities using Equation~\ref{density_3} and assuming the line-of-sight thickness $D$ between $2.9~\text{Mm}$ and $40.7~\text{Mm}$ for regions~A and~B. Although the electron densities for the two regions are quite similar during the observation, region A shows a slightly higher density, which is expected as it is closest to the photosphere and consequently influenced by the material just above the bright loops. We performed an error propagation analysis to determine the uncertainty in our calculation of the column density and hence the electron density, finding this relative error to be about 11\%, and to be dominated by the observational error of Q'. 
An additional and ill-defined uncertainty is due to the assumption of a value for the column depth.

This geometrical uncertainty does not affect the mass $M$ estimated via Equation~\ref{eq:scat_mass}, using the excess flux $F_S$ observed from each source region.
From the first entry in Table~\ref{tab:obs}, we find that $F_s/F_\odot = 6.7 \times 10^{-9}$ for Box A at 16:19~UT. This corresponds to an electron number $4 \times 10^{38}$ and a (total) mass $8 \times 10^{14}$~g.
This mass estimate is interesting as it is a simple to measure the gravitational potential energy of the scattering source, here roughly $5 \times 10^{28}$~erg.
For further discussion see Section~\ref{sec:interp}.

\subsection{X-ray and microwave observations}

The polarized continuum observed by HMI relates directly to the free-free and free-bound continua observed in the microwave and X-ray bands.
The same electrons contribute both to the emission measure ($\propto n_e^2$) and to the polarized flux ($\propto n_e$).
The X-ray continuum also has significant free-bound and bound-bound contributions, whereas the microwaves do not.
Each can serve to estimate the value of $n_e n_i \eta V$, and all three wavelength ranges should agree for a hot isothermal source according to 
Equation \ref{eq:free-free} \citep[e.g.][]{1972SoPh...23..155H}:

\begin{equation}\label{eq:free-free}
    f_\nu \propto (n_e n_i V) \Lambda(n_e, T) \cdot \frac{e^{-h\nu/kT}}{\sqrt{T}}\ \ ,
\end{equation}
where $f_\nu$ is the spectral flux in erg~(cm$^2$~s~Hz)$^{-1}$,
including the slowly varying Coulomb logarithm $\Lambda(n_e, T)$.
Note that the microwave flux has only a weak dependence upon the source temperature, but that it may have low-frequency cutoffs due to opacity or the Razin-Tsytovich efffect \citep[e.g.][]{2013A&ARv..21...58K}; furthermore the relatively weak thermal microwave continuum also may have to  compete with other emission mechanisms such as gyrosynchrotron or plasma radiations.

\begin{figure}[htbp]
\centering
\includegraphics[width=0.6\textwidth]{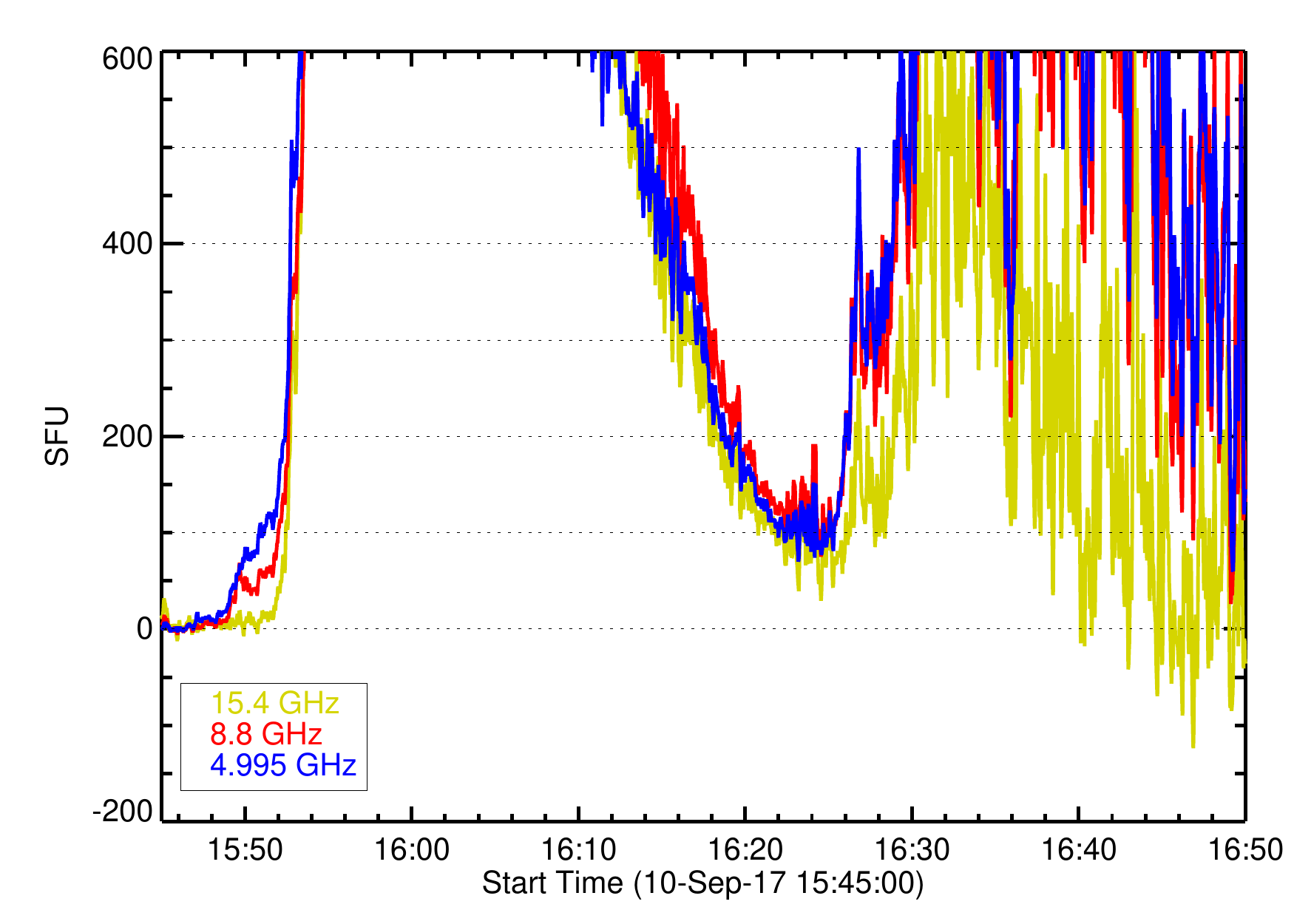}
          \caption{\textit{Single-frequency observations of SOL2017-09-10 from RSTN/San Vito (4.995, 8.8, and 15.4~GHz), with pre-event background levels subtracted. The minimum coincides approximately with the Box A snapshot time. The radio spectral irradiances are measured in standard solar flux units, 10$^{-22}$~W~(m$^2$ Hz)$^{-1}$.} 
      }
\label{fig:SOL2017-09-10_RSTN}
\end{figure}

Figure~\ref{fig:SOL2017-09-10_RSTN} shows single-frequency data from the San Vito site of the Radio Solar Telescope Network (RSTN).
Based on Equation~\ref{eq:free-free}, we expect to see a long-wavelength extension of the X-ray and white light continua through the IR, mm-wave, and longer radio wavelengths.
This extension should have only the weak frequency dependence of the Gaunt factor, and so $f_\nu \approx {\mathrm const.}$ -- a flat spectrum.
What we see in the Figure is different from this expectation, mainly because of gyrosynchrotron emission.
The major time-series peak prior to 16:20~UT is likely to be from this mechanism, and the emission after about 16:25~UT may also come from a competing source.
We interpret the minimum at around 16:20-25~UT in terms of the thermal signature of the hot plasma detected by RHESSI, but only as an upper limit.
The spectral character of the minimum-flux epoch is consistent with optically thin free-free emission (Equation~\ref{eq:free-free}, which could reflect the long-wavelength extension of the RHESSI hot plasma and the polarized white-light continuum.
The microwave flux does indeed have the right magnitude for the free-free interpretation, but this is just a consistency check.

The two boxes A and B introduced in Figure~\ref{HMI-AIA} have been chosen to coincide with the main X-ray sources seen by RHESSI.
We defer the details of the RHESSI measurements to Section~\ref{sec:rhessi} but report the result for the thermal emission measure of the source in Box A here.
From the thermal component of the spectral fits we can derive a volumetric emission measure $n_e n_i \eta V$~cm$^{-3}$ for the hot component in terms of the volumetric filling factor $\eta$.
We compare this with the same quantity derived from the whole-Sun fluxes observed by GOES/XRS.
Finally we can also interpret the RSTN\footnote{\url{http://www.ngdc.noaa.gov/stp/space-weather/solar-data/solar-features/solar-radio/rstn-1-second/}} whole-Sun microwave observations for this parameter, noting that the lack of temperature senstivity for the microwave thermal component makes it possible to check for low-temperature flare sources not visible in soft X-rays \citep[e.g.,][]{2016ApJ...819L..30P}.
Table~\ref{tab:em} lists all three of these estimates.

\begin{table}[h]
\smallskip
\caption{Volumetric Emission Measures\label{tab:em}}
\begin{center}
    \begin{tabular}{ l r}
    \hline\noalign{\smallskip}
    Source & Value (cgs) \\
    \hline\noalign{\smallskip}
    RHESSI &  $(3.0  \pm 0.3) \times 10^{50} $ \\
    GOES-15/XRS &  $(3.0 \pm 0.1) \times 10^{50}$  \\ 
    RSTN/San Vito & $<(3.3  \pm 0.5) \times 10^{50} $  \\
    \hline\noalign{\smallskip}
    \end{tabular}                    
\end{center}
\end{table}

\begin{figure}[htbp]
\centering
\includegraphics[width=0.49\textwidth]{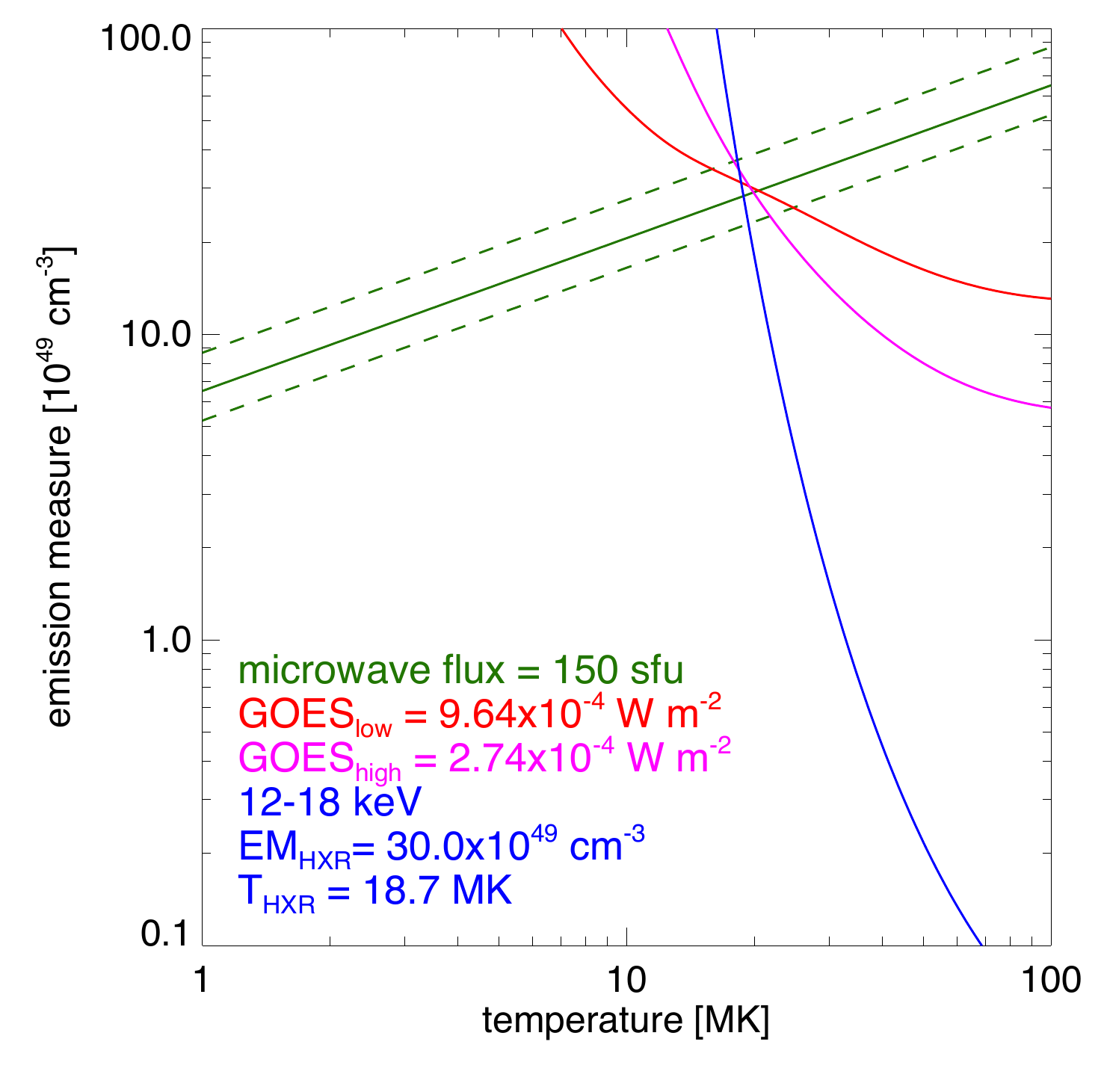}
\includegraphics[width=0.49\textwidth]{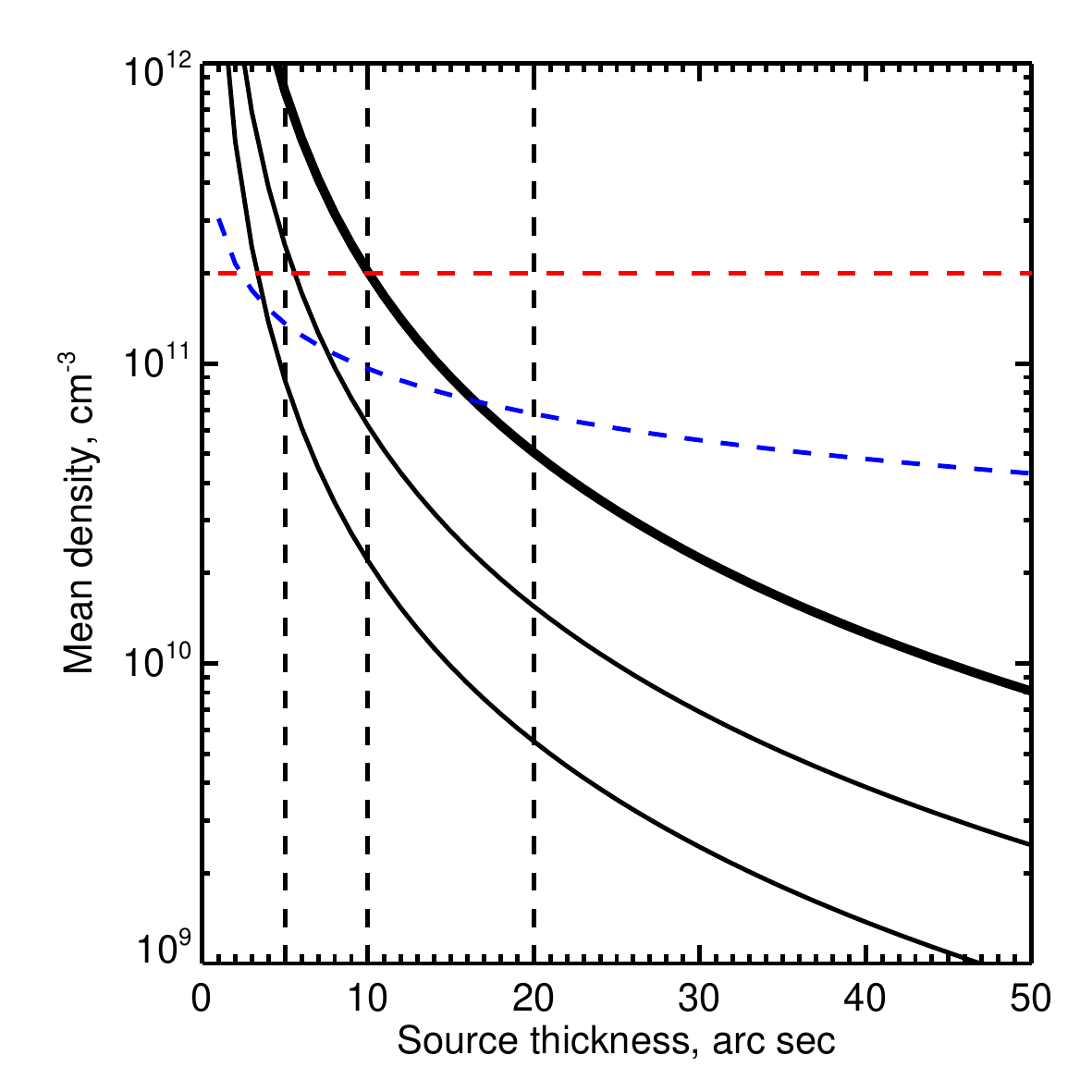}
          \caption{\textit{Left: the ``loci plot'' comparing the microwave emission measure with X-rays.
          The GOES and RHESSI exclusion lines agree well with the RHESSI 12-18~keV flux, and match the short-wavelength extrapolation of the microwave thermal flux.
          Right (Section~\ref{sec:results}): constraints on the mean source density for the polarization source, combining HMI observations with X-ray and microwave data.
          The colored lines at top are limits imposed by the Razin-Tsytovich effect (red) and by free-free opacity (blue), in the absence of a long-wavelength spectral break \citep[e.g.,][]{1996AIPC..374..416B}.
          The heavy black line corresponds to filling factory unity, with the thin lines for values of 0.3 and 0.1.
          The vertical dashed lines are the Box~A width times 0.5, 1, and 2.} 
      }
\label{fig:loci_plot}
\end{figure}

This agreement between microwave and X-ray emission measures confirms the identification of the sources and precludes the presence of substantial flare emission at low temperatures, according to Equation~\ref{eq:free-free}, this would destroy the agreement between microwave and X-ray emission-measure estimates displayed in the ``loci plot'' of Figure~\ref{fig:loci_plot}.

The agreement of the HMI estimate of column density $N_e$ with the radio and X-ray values also allows us, uniquely for this flare, to estimate the line-of-sight thickness of the source $\eta D$ in terms of the observed volumetric filling factor EM and the total electron number ${\mathcal N}$ as derived in Section~\ref{sec:cd}.
The emission measure implies an electron content
${\mathcal N}' = n_e' V' =  \sqrt{EM \cdot \eta D A}$, which we can combine with the true electron number to give the estimate for the source thickness
\begin{equation}\label{eq:thickness}
\eta D = \frac{{\mathcal N}^2}{A\cdot EM}\ ,\end{equation}
based on simple estimates of the source area $A$ and depending on the volumetric filling factor $\eta$.

\subsection{RHESSI imaging}\label{sec:rhessi}

The final information needed for characterizing the polarization sources comes from the RHESSI imaging, from which we can obtain estimates of the projected area of the joint X-ray/Thomson-scattering/thermal-microwave source identified by the agreement in Table~\ref{tab:em}.

At the time of the observations of the September 10 flare, the RHESSI spectrometer had been operating in space for more than 16 years and only 4 of the 9 detectors were functioning. 
Even for these four detectors (detectors 1, 3, 6, and 8), radiation damage had substantially affected the spectral resolution and effective area. 
Because radiation damage affects each detector differently, we performed spectral fitting for each detector separately and then compare the derived parameters to judge their consistency. 

The RHESSI observations ended at about 16:17:40~UT when the RHESSI spacecraft entered the South Atlantic Anomaly. We selected the interval from 16:16:30 to 16:17:20~UT for our comparison with the HMI observations taken at 16:19~UT. The orbit-night background from 15:49 to 15:52~UT has been subtracted before the standard RHESSI spectral fitting of a thermal component plus a non-thermal component was done.  The thermal fits dominate the total count spectrum for energies below 25~keV, so we take the 12-18 keV images shown in Figure~1 to represent the thermal sources. The derived temperatures from the four different detectors range from 18.2 to 19.3~MK in good agreement. The corresponding emission measures vary from 2.6 to 3.5 $\times 10^{50}$ cm$^{-3}$, with the  average value of 3.3$\pm$0.5$\times 10^{50}$ cm$^{-3}$ as quoted above in Table~\ref{tab:em}. 
The temperature and emission measure derived from GOES-15 are 19.1~MK  and 3.0$\pm$0.1$\times 10^{50}$ cm$^{-3}$, respectively, in surprisingly good agreement with RHESSI, considering the effects of radiation damage to the effective area of the RHESSI detectors. 

The RHESSI parameters derived above refer to the total counting rates.
We can also use imaging spectroscopy to estimate temperature and emission measure for the RHESSI source contained within Box~A \citep[e.g.][]{2002SoPh..210..229K}. The fluxes from within Box~A are derived from CLEAN images at 1~keV resolution, which were reconstructed in the same way as the images shown in Figure 1. Using the energy range 13--18~keV gave parameters [21.0$\pm$1.1 MK, 1.2$\pm$0.4$\times 10^{50}$ cm$^{-3}$]. Here the errors have been derived by adding Gaussian random noise at 10\% rms in a Monte Carlo procedure; this gave parameters [19.2--22.6~MK, and 0.6--2.0 $\times 10^{50}$ cm$^{-3}$]. 
The derived temperature of the main source is thus slightly hotter than the flare-averaged value. Considering the uncertainties the difference is of minor significance. The emission measure of the bright source is about a factor of 3 smaller, indicating that the fainter sources outside Box A  cannot be neglected and indeed affect the flare-averaged fit parameters. 

In the final step of the RHESSI analysis, we estimate the source area of the thermal hard X-ray source. We use CLEAN images to estimate the source size of the main compact RHESSI source (source A). From the RHESSI clean image shown in Figure~1 (top), we derive a FWHM source size of the compact part of the source of 15.8$''$. The image is reconstructed with a CLEAN beam of 12.2$''$. Hence, we get a deconvolved source size of 10.0$''$.

\subsection{The Box~B interval}

RHESSI exited the South Atlantic Anomaly (SAA) around 16:39~UT; the second interval (Box~B) was fully observed by RHESSI with both attenuators still inserted. 
The same analysis is applied to the later interval as has been done for the earlier interval using the orbit-night background from 16:55 to 16:58~UT. The non-thermal fit is marginal at this time, with very steep spectral slopes ($\approx 8$). 
The temperature ranges from 16.0 to 17.7 MK with an average value of 17.0$\pm$0.6 MK; and the emission measures vary between 0.7 to 1.3 $\times10^{50}$cm$^{-3}$, with an average value of $1.0\pm0.2\times10^{50}$cm$^{-3}$.  
GOES at this time gives [T, EM] = [15.5~MK, 1.3$ \times 10^{50}$cm$^{-3}$], again in good agreement with RHESSI.
Imaging-spectroscopy fits for the main compact source give higher temperatures (22.8$\pm$1.1 MK). As the temperature of the main source is higher than the flare-averaged temperature, a significantly lower emission measure of $0.10\pm0.04\times10^{50}$cm$^{-3}$ of the compact source already produces a good fit with the observed hard X-ray flux in the box B. As extreme values for the emission measure we found $0.04\times10^{50}$cm$^{-3}$ and $0.3\times10^{50}$cm$^{-3}$.

The CLEAN image shown in Figure~1 has an area with the 50\% contour level of 18.5$''$ $\times$18.5$''$, giving a deconvolved source diameter of 14.0$''$. Compared to the earlier interval around 16:19~UT, the source area apparently had roughly doubled in area.   

\section{Results}\label{sec:results}

With these data we can now incorporate the polarization observations into a description of the RHESSI soft X-ray source.
The essential augmentation or our knowledge is twofold: we can directly estimate the number of electrons, and hence the mass, of the scattering source; in addition that we can determine the product $\eta D$ (Equation~\ref{eq:thickness}).

\subsection{Electron number, mass, and energy}\label{sec:numbers}

As described above, the measurement of scattered flux directly determines the total number of free electrons, with the plausible assumptions of plane-of-the-sky location and nominal limb-darkened solar radiance \citep{2021ApJ...923..276S}.
We find $4\ \times\ 10^{38}$ for the total electron number.
This in turn determines the coronal mass at the time of observation, plus its gravitational potential energy.
The values we have derived for the Box~A snapshot are $8 \times 10^{14}$~g and $5 \times 10^{28}$~erg, respectively.

\subsection{Constraints on density and filling factor}

The right panel of Figure~\ref{fig:loci_plot} summarizes the constraints imposed by the observed value of $\eta D = 4 \times 10^{9}$~cm.
To this we have added constraints from the Razin-Tsytovich effect and from free-free opacity \citep{1996AIPC..374..416B}.
These limits are based on the assumption that the RSTN three-point spectrum at the Box~A epoch (Figure~\ref{fig:SOL2017-09-10_RSTN}) does not show a spectral break.
The free-free limit definitely restricts the density estimate for unity filling factor,
shown in the Figure as the thick solid line (the depth-dependence of mean density for $\eta$~=~1).
For some part of the range of possible source thicknesses, we can exclude this limit.
We would expect a filling factor below unity in any case from the fine structure in the developing arcade, but there is no real constraint from the microwave or X-ray imaging.
From the diagram we conclude that the probable value for the mean source density does not exceed $2 \times 10^{10}$~cm$^{-3}$ for line-of-sight depths comparable to the dimension of the integration box.

\subsection{Interpretation}\label{sec:interp}

This remarkable solar flare (SOL2017-09-10) has allowed us to consider in detail the information about hot coronal emission sources in a major flare.
These sources consist of hot plasma, and we have shown for the first time that observations of linear polarization in the optical continuum match the extrapolated soft X-ray thermal emission spectrum right down to the microwave range.
This convincingly establishes that flares do not require the injection into the corona of cool material (below about 1~MK), which would readily be detectable in the Paschen continuum and in the microwave free-free emission in the Rayleigh-Jeans spectrum \citep[e.g.][]{2016ApJ...819L..30P}.
The absence of few-MK temperatures is also consistent with the AIA-derived emission measure distributions for this event, as reported by \cite{2018ApJ...854..122W}.
The polarization brightness is directly proportional to the mass of the hot coronal cloud; this information cannot be obtained otherwise except in model-dependent ways.
The coronal mass, $8 \times 10^{14}$~g, estimated some 30 minutes after the onset of the flare's impulsive phase, suggest that the magnetically trapped mass can rival that of the most massive CME, often quoted as about 10$^{16}$~g.
The potential energy estimate, however, is less than a percent of total energy estimates for such an event, either flare or CME \citep[e.g.][]{2012ApJ...759...71E}.

The standard picture of the development of a major eruptive flare envisions a ``magnetic explosion'' \citep[eg][]{2001ApJ...552..833M} creating a CME, followed by an extended period of magnetic reconnection that forms the loop prominence system finally seen in chromospheric emission lines.
The plasma that fills these loops, according to this well-accepted picture, originated via chromospheric ``evaporation'' and appear as the hot cloud our observations show and as depicted in many cartoon descriptions.
This hot mass has been driven into the corona and heated, although the details of how this happens remain unclear. 
Our data are consistent with the idea that the heating can be identified with particle acceleration leading to plasma $\beta$ of order unity \citep{2010ApJ...714.1108K}.
The standard picture, dating back at least to \cite{1964ApJ...140..746B}, involves hot mass in larger and larger loops cooling sequentially, implying that the instantaneous mass we infer at the epoch of Box~A is a lower limit to the total mass for the entire flare process.
After cooling, the ionized plasma recombines and may become optically thick, radiating strongly via collisional processes \citep{2018ApJ...867..134J} at densities much higher than we infer for the hot plasma.

\section{Summary}
The successful measurement of linear polarization in the optical continuum has allowed us to make direct estimates of the electron content of the flare's coronal plasma, and hence its mass.
The information provided by this new tool is in agreement with previous observations of classical loop prominence systems \citep{1964ApJ...140..746B,2007SoPh..246..393S,1992PASJ...44...55H}.
Essentially we have extended the familiar methods of K-corona studies to the bottom of the corona \citep{2021ApJ...923..276S}.

The work reported here made serendipitous use of HMI data, which were never intended for studying off-limb sources
and are limited in performance, for this purpose, by design factors in the optics and the spectral bandwidth.
In particular, we have not attempted to use these data to analyze either the image structure of the polarization or its development in time.
Both of these would contain novel information about the evolution of the excess coronal mass created by the flare.
Our simple mass estimate is of special interest now because of the remarkable EOVSA microwave observations of this flare (SOL2017-09-10), which reveal important processes outside the regions of the EUV loop/cusp/current-sheet geometry \citep{2020ApJ...895L..50C,2022Natur.606..674F}.
Our multi-wavelength analysis does suggest that the polarization source can be identified with its counterparts in the soft X-rays and microwaves at high electron temperatures, and can disambiguate these observations in terms of electron density and filling factor.
This just gives us a taste of what may be learned by better-optimized observations in the future.
Better imaging polarimetry would allow for a far deeper investigation of the coronal structure of a 
suitable flare  and of its evolution \citep[\textit{e.g.} DKIST,][]{2021SoPh..296...70R}.

\acknowledgments

J.C.G.G. wish to acknowledge the SolarALMA project, which has received funding from the European Research Council (ERC) under the European Union’s Horizon 2020 research and innovation programme (grant agreement No. 682462), and by the Research Council of Norway through its Centres of Excellence scheme, project number 262622.
H.S.H. thanks the University of Glasgow for hospitality during the completion of this work.

\bibliography{biblio}
\bibliographystyle{aasjournal}

\end{document}